\documentclass[11pt]{article}
\usepackage{graphicx}

\newcommand{\BABARPubYear}    {00}

\newcommand{\BABARConfNumber} {16}
\newcommand{\SLACPubNumber} {8538}

\usepackage{relsize}
\def\babar{\mbox{\slshape B\kern-0.1em{\smaller A}\kern-0.1em
    B\kern-0.1em{\smaller A\kern-0.2em R}}}

\def\pip   {\ensuremath{\pi^+}}
\def\pim   {\ensuremath{\pi^-}}

\def\Kbar  {\kern 0.2em\overline{\kern -0.2em K}{}}

\def\Kp    {\ensuremath{K^+}}

\def\Kstarz  {\ensuremath{K^{*0}}}

\def\Kzb   {\ensuremath{\Kbar^0}}
\def\KzKzb {\ensuremath{K^0 \kern -0.16em \Kzb}}

\def\Dbar  {\kern 0.2em\overline{\kern -0.2em D}{}}

\def\Dzb   {\ensuremath{\Dbar^0}}
\def\DzDzb {\ensuremath{D^0 {\kern -0.16em \Dzb}}}

\def\Bz    {\ensuremath{B^0}}

\def\Bbar  {\kern 0.18em\overline{\kern -0.18em B}{}}

\def\Bzb   {\ensuremath{\Bbar^0}}
\def\Bu    {\ensuremath{B^+}}

\def\BB    {\ensuremath{B\Bbar}} 
\def\BzBzb {\ensuremath{B^0 {\kern -0.16em \Bzb}}}

\def\jpsi  {\ensuremath{{J\mskip -3mu/\mskip -2mu\psi\mskip 2mu}}} 
\def\psitwos {\ensuremath{\psi{(2S)}}}
\mathchardef\Upsilon="7107
\def\Y#1S{\ensuremath{\Upsilon{(#1S)}}}% no space before {...}!

\def\FourS {\Y4S}

\mathchardef\Deltares="7101
\mathchardef\Xi="7104
\mathchardef\Lambda="7103
\mathchardef\Sigma="7106
\mathchardef\Omega="710A
\def\Deltabar   {\kern 0.25em\overline{\kern -0.25em \Deltares}{}}
\def\Lbar {\kern 0.2em\overline{\kern -0.2em\Lambda\kern 0.05em}\kern-0.05em{}}
\def\Sigbar{\kern 0.2em\overline{\kern -0.2em \Sigma}{}}
\def\Xibar{\kern 0.2em\overline{\kern -0.2em \Xi}{}}
\def\Obar{\kern 0.2em\overline{\kern -0.2em \Omega}{}}
\def\Nbar{\kern 0.2em\overline{\kern -0.2em N}{}}
\def\Xbar{\kern 0.2em\overline{\kern -0.2em X}{}}

\def\BR{{\ensuremath{\cal B}}}

\def\mes        {\mbox{$m_{\rm ES}$}}

\def\ev   {\ensuremath{\rm \,e\kern -0.08em V}}
\def\kev  {\ensuremath{\rm \,ke\kern -0.08em V}} 
\def\mev  {\ensuremath{\rm \,Me\kern -0.08em V}} 
\def\gev  {\ensuremath{\rm \,Ge\kern -0.08em V}} 
\def\gevc {\ensuremath{{\rm \,Ge\kern -0.08em V\!/}c}} 
\def\tev  {\ensuremath{\rm \,Te\kern -0.08em V}}
\def\mevc {\ensuremath{{\rm \,Me\kern -0.08em V\!/}c}} 
\def\gevcc{\ensuremath{{\rm \,Ge\kern -0.08em V\!/}c^2}} 
\def\mevcc{\ensuremath{{\rm \,Me\kern -0.08em V\!/}c^2}}

\def\cm   {\ensuremath{\rm \,cm}}

\def\invfb   {\ensuremath{\mbox{\,fb}^{-1}}}
\def\mus  {\ensuremath{\rm \,\mus}}

\def\mus        {\ensuremath{\,\mu{\rm s}}}    %% microsecond
\def\gsim{{~\raise.15em\hbox{$>$}\kern-.85em
          \lower.35em\hbox{$\sim$}~}}
\def\lsim{{~\raise.15em\hbox{$<$}\kern-.85em
          \lower.35em\hbox{$\sim$}~}}

\def\to                 {\ensuremath{\rightarrow}}

\def\pep2{PEP-II}

\newcommand{\dedx}{\ensuremath{\mathrm{d}\hspace{-0.1em}E/\mathrm{d}x}}

\newcommand{\eqref}[1]{Eq.~(\ref{eq:#1})}

\newcommand{\epjc}      [1]  {{Eur.\ Phys.\ Jour.\ C~{\bf #1}}}

\def\geant      {\mbox{\tt Geant321}}

\def\jetset74   {\mbox{\tt Jetset \hspace{-0.5em}7.\hspace{-0.2em}4}}

\long\def\inst#1{\par\nobreak\kern 4pt\nobreak
    {\it #1}\par\vskip 10pt plus 3pt minus 3pt}

\begin{document}
{\pagestyle{empty}

\begin{flushright}
\babar-CONF-\BABARPubYear/\BABARConfNumber \\
SLAC-PUB-\SLACPubNumber
\end{flushright}

\par\vskip 3cm

% Title of the paper
\begin{center}
\Large \bf Search for 
%{\boldmath $ B \rightarrow K(K\ast) \ \ell^+ \ell^-$}
{\boldmath $ \Bu \rightarrow \Kp     \ \ell^+ \ell^-$} and
{\boldmath $ \Bz \rightarrow \Kstarz \ \ell^+ \ell^-$}  
\end{center}
\bigskip

\begin{center}
\large The \babar\ Collaboration\\
\mbox{ }\\
July 25, 2000
\end{center}
\bigskip \bigskip

% Abstract
\begin{center}
\large \bf Abstract
\end{center}
Using a sample of $3.7 \times 10^{6}$ $\Upsilon(4S) \rightarrow \BB$ events
collected with the \babar\ detector at the \pep2\ storage ring, we search
for the electroweak penguin decays $B^+ \rightarrow K^+ e^+ e^-$,
$B^+ \rightarrow K^+ \mu^+ \mu^-$, $B^0 \rightarrow \Kstarz \ e^+ e^-$, and
$B^0 \rightarrow \Kstarz \ \mu^+ \mu^-$. 
We observe no significant signals for these modes and 
set preliminary 90\% C.L. upper limits of
\begin{eqnarray*}    
\BR(B^+ \rightarrow K^+ e^+ e^-)  & <  & 12.5 \times 10^{-6}, \\
\BR(B^+ \rightarrow K^+ \mu^+ \mu^-) &  < & \phantom{2}8.3 \times 10^{-6}, \\
\BR(B^0 \rightarrow \Kstarz e^+ e^-) & < &  24.1  \times  10^{-6}, \\
\BR(B^0 \rightarrow \Kstarz \mu^+ \mu^-) & < & 24.5 \times 10^{-6}. \\
\end{eqnarray*}     

\vfill
\centerline
{Submitted to the XXX$^{th}$ International Conference on High Energy Physics, Osaka, Japan.}  
\newpage
}

% Input author list file
\begin{center}
\small

The \babar\ Collaboration
\bigskip

%% author list as of 24-Jul-2000 (580 authors) % hand edited removal of street addresses by dbmacf
%% 8-4-00: fixed addresses for UC campuses; added Sh. Rahatlou to UCSD: dbmacf
B.~Aubert,
A.~Boucham,
D.~Boutigny,
I.~De Bonis,
J.~Favier,
J.-M.~Gaillard,
F.~Galeazzi,
A.~Jeremie,
Y.~Karyotakis,
J.~P.~Lees,
P.~Robbe,
V.~Tisserand,
K.~Zachariadou
\inst{Lab de Phys.\ des Particules, F-74941 Annecy-le-Vieux, CEDEX, France}
% ...

\end{center}\newpage

% reset footnote counter
\setcounter{footnote}{0}

% The body of the paper starts here
\section{Introduction}
\label{sec:Introduction}

The rare decays $B\to K\ell^+\ell^-$ and $B\to K^*\ell^+\ell^-$,
where $\ell$ is either an electron or muon, are highly
suppressed in the Standard Model and are expected to occur via 
electroweak penguin processes. The internal loops in such
processes lead to a $b\to s$ transition via an effective
flavor-changing neutral current interaction. 
Standard Model 
predictions for these decays are dependent to some extent
on the modeling of the strong interactions in the hadronic
$B\to K^{(*)}$ transition, 
but calculations 
indicate that
$\BR(B\to K\ell^+\ell^-)\approx 6\times 10^{-7}$,
while $\BR(B\to K^*\ell^+\ell^-)\approx 2\times 10^{-6}$~\cite{bib:ali1}.
These processes provide a possible window into physics beyond the Standard Model,
since new, heavy particles such as those predicted by
SUSY models can also enter the loops,  
resulting in significant changes to both the decay 
rates and kinematic distributions~\cite{bib:buchalla}. 

Experimentally, the small expected rates make searches
for these modes difficult. In addition, backgrounds from 
tree-level processes, such as $B\to J/\psi K^{(*)}$, with
$J/\psi\to\ell^+\ell^-$, complicate the analyses.
Searches from CDF~\cite{bib:CDF} and CLEO~\cite{bib:CLEO1} have so far yielded 
only upper limits, although in the case of $B\to K^*\mu^+\mu^-$ the limit
is only about a factor of two above the Standard Model prediction. 

In this paper, we report the results of a preliminary analysis to investigate
the backgrounds and the ability of the \babar\ detector to reject them. 
We have analyzed an on-resonance data sample of 3.2\invfb,
representing about a third of the current \babar\ $\Upsilon(4S)$
integrated luminosity.

As in many other searches for very rare decays, there are important
issues that we face in designing event selection criteria in
a manner that does not bias the determination of the signal yield,
or, at worst, create a false apparent 
signal in a small statistics event sample.
The main goal of our study is to test the performance of 
a ``blind'' analysis in which 
the event selection is optimized without use of the signal
or sideband regions in the data. The events in these regions 
are therefore selected without
bias, and the sidebands can be used
for background estimation. The decays $\Bu \to J/\psi K^+$ and
$B^0\to J/\psi K^{*0}$, which have the same topology as the
signal, provide a useful control sample to compare data with Monte Carlo
simulation. These events can be used to validate the efficiencies
predicted by the full \geant ~Monte Carlo simulation.
An additional control sample is provided by events with 
$K^{(*)}e^{\pm}\mu^{\mp}$
candidates. Such events monitor the level of combinatorial
backgrounds present in the sample.  
This analysis provides a basis for
extending the measurement to the much larger data samples that \babar\
will obtain in the future. 
The 3.2 \invfb\ data sample is comparable in size, however, to that used
in an earlier CLEO analysis, and we are able to obtain results of
comparable sensitivity for the $B^+ \rightarrow K^+ \ell^+ \ell^-$ modes.  

We analyze four decay modes: 
$B^+\to K^+ e^+e^-$, $B^+\to K^+ \mu^+\mu^-$, 
$B^0\to K^{*0} e^+e^-$, and $B^0\to K^{*0} \mu^+\mu^-$.
The $K^{*0}$ is reconstructed in the $K^+\pi^-$ final
state. In each case, we include the charge conjugate mode as well.

\section{The \babar\ detector and data sample}
\label{sec:babar}
The \babar\ detector is described in detail elsewhere~\cite{bib:babar}. All components
of the detector are used for this study. Of particular importance for this
analysis are the charged tracking system and the detectors used for
particle identification. At radii between about 3 cm and 14\cm,
charged tracks are measured with high precision in a five-layer
silicon vertex tracker (SVT). This device comprises 52 modules built
from double-sided AC-coupled silicon strip detectors. Tracking beyond
the SVT is provided by the 40 layer drift chamber, which extends from 23.6\cm\
to 80.9\cm.  
Just outside the drift chamber is the DIRC, which
is a Cherenkov ring-imaging particle identification system. Cherenkov light
is produced by charged tracks as they pass through an array of 144 
five-meter-long fused silica quartz bars; the light is transmitted to
the ends of the bars by total internal reflection, preserving the 
information on the angle of the light emission with respect to the
track direction. Electrons are detected using
an electromagnetic calorimeter comprising 6580 thallium-doped CsI
crystals. Muons are detected in the Instrumented Flux Return (IFR), 
in which resistive plate chambers (RPCs) are interleaved
with the iron plates of the flux return.

The data used in the analysis were collected with the \babar\ detector
at the \pep2\ storage ring.  We analyze a subsample of this data consisting
of 3.2\invfb\ taken on the $\Upsilon(4S)$ resonance and 1.2\invfb\ at
a slightly lower energy to obtain a pure continuum sample. 
%The $K^{(*)}e^{\pm}\mu^{\mp}$ control
%sample was obtained from the larger 8.0 fb$^{-1}$ on resonance data set.
For the purpose of optimizing the
event selection, the following simulated event samples were used: 
\begin{itemize}
\item 4.2 million generic $\FourS\to\BB$
events generated using the \geant~package, 
\item 4.2 million continuum $c\bar c$ events (\geant),
\item 4.8 million continuum $u\bar u$, $d\bar d$, and $s\bar s$ events
(\geant), 
\item 50\invfb\ of a fast, parametrized Monte Carlo simulation
with the correct mixture
of \BB\ and continuum events,
\item 5,000 events for each signal mode (\geant).
\end{itemize}

\section{Kinematic properties of signal events}

The properties of $\Upsilon(4S)\to \BB$ events containing a
$B\to K^{(*)}\ell^+\ell^-$
decay are modeled with a full \geant~Monte Carlo simulation using the event
generator EvtGen~\cite{bib:EVTGENref}. We have implemented a generator based on the
matrix elements computed in Ref.~\cite{bib:ali1}. Although there are significant
uncertainties due to strong interaction effects, the matrix elements can be
rigorously parametrized in terms of form factors, which are Lorentz-invariant functions
of $q^2=M^2_{\ell^+\ell^-}$. 

In the regions of $q^2$ near the \jpsi\ and \psitwos\ resonances, interference
occurs between the penguin amplitude and that for charmonium production, since
the \jpsi\ and \psitwos\ can decay into $\ell^+\ell^-$.
This interference enhances the
rate on the low-mass side of the resonance and suppresses it on the high side.  These
effects are modeled in Ref.~\cite{bib:ali1}, but not in our Monte Carlo generator.
However, since
$B\to J/\psi K^{(*)}$ and $B\to\psitwos K^{(*)}$ are major backgrounds in our analysis,
we eliminate most of these mass regions, as described in the following section. 
Small residual interference effects are expected for masses close to our exclusion regions,
but we have ignored these in the present study. Further enhancements in the rate,
along with related interference effects, are expected
from charmonium resonances with masses above the \psitwos. These effects are relatively
small compared to the expected electroweak penguin contribution, and they are
ignored in this study.

\begin{figure}[!htb]
\begin{center}
\includegraphics[height=5.5in]{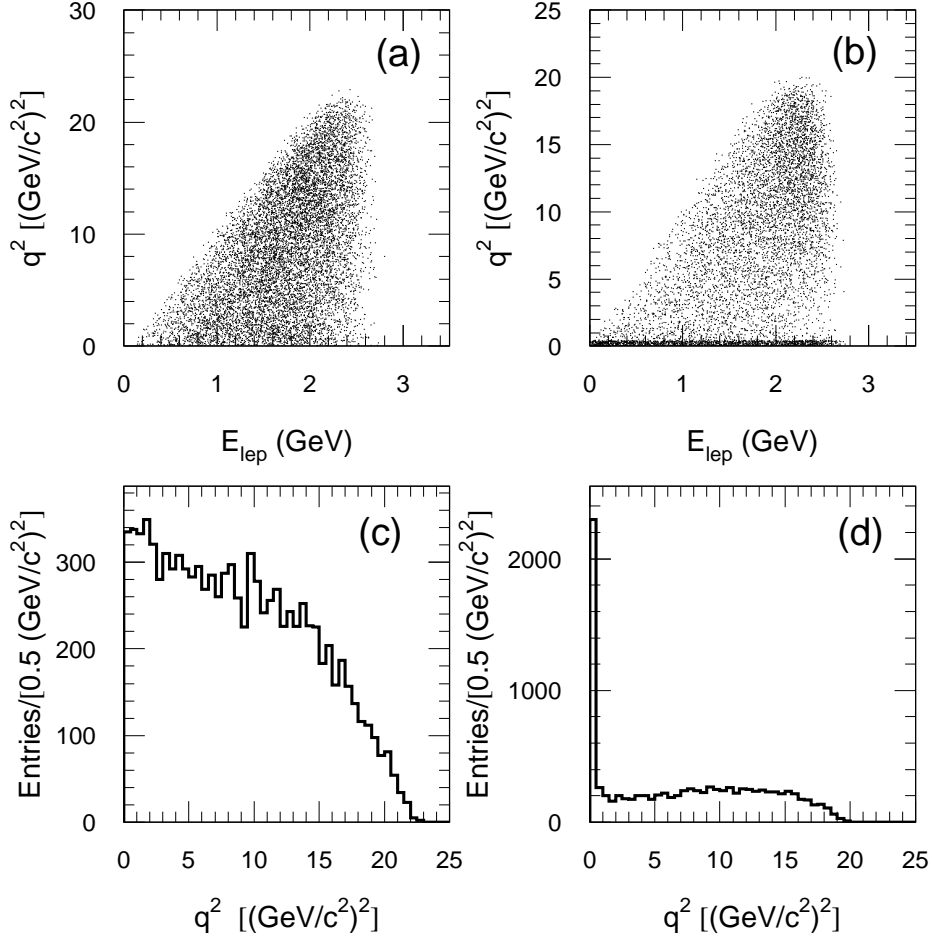}
\caption{Monte Carlo simulated distributions for signal processes
(a) Dalitz plot for $B\to Ke^+e^-$, (b) Dalitz plot for $B\to K^* e^+e^-$, (c) $q^2=M^2_{e^+ e^-}$ distribution
for $B\to Ke^+e^-$, and (d) $q^2$ distribution for $B\to K^* e^+ e^-$.
}
\label{fig:Dalitz}
\end{center}
\end{figure}

The Dalitz plot and $q^2$ distributions resulting from
these matrix elements are shown in Fig.~\ref{fig:Dalitz}. For Dalitz
plot variables
we have chosen $q^2$ and $E_{\rm lep}$, where the lepton energy is given in the $B$ meson
rest frame, and the positive lepton is selected for the case $B^+\to K^+ \ell^+\ell^-$ and $B^0 \to K^{\ast}\ell^+ \ell^-$ (the negative lepton is selected for the charge-conjugate states).
For $B\to K\ell^+\ell^-$, the $q^2$ 
distribution peaks at low values, corresponding to rapid $K$ recoil in the $B$
rest frame. For $B\to K^*\ell^+\ell^-$ the $q^2$ distribution peaks at large values,
corresponding to slow $K^*$ recoil in the $B$ rest frame, but there is
a pole in the amplitude at $q^2=0$, where the photon becomes real. The decay
$B\to K\ell^+\ell^-$ does not have a pole at $q^2=0$ because $B\to K\gamma$ 
is forbidden by angular momentum conservation.

\section{Overview of the analysis method}
\label{sec:Analysis}

The analysis method exploits the strong signature and kinematic constraints
for the $B\to K^{(*)} \ell^+\ell^-$ decay
and the broad capabilities of the 
\babar\ detector. Lepton and kaon identification,
good momentum measurement, and the constraint associated with the known beam energies
are essential for the search. 
Because even low-level backgrounds can be problematic, we
have to be concerned about a possible ``cocktail'' of small contributions. 
Some of the most important sources are: 
\begin{itemize}
\item $B\to \jpsi K^{(*)}$ or $B\to \psitwos K^{(*)}$ with \jpsi\ or
$\psitwos \to\ell^+\ell^-$. If one of the leptons radiates a photon, the
mass of the dilepton system can fall below the resonance veto region. These events
must be rejected with high efficiency, since they have precisely the same topology
as signal events.  
\item Combinatorial backgrounds from random leptons and kaons in \BB\ and continuum
processes ($e^+e^-\to q\bar q$, where $q=u$, $d$, $s$, or $c$). In general, the lepton
candidates arise
from semileptonic decays, although leptons from photon conversions or from hadrons faking
a lepton signature also contribute at a low level.
\item Certain backgrounds can peak in the signal region, but these are quite rare and
easily rejected. An example is $\Bu\to\Dzb\pip$
with $\Dzb\to\Kp\pim$. If both pions are
misidentified as muons, this process can be confused with the signal. This type of three-body
background is easily rejected with suitable reassignment of particle
identification hypotheses and
a veto on the $D$ mass region.
\end{itemize}

Our goal in the present search is to gain an understanding of background levels and
distributions and to set limits. For this purpose,
we assume that each event in the signal region is potentially due to the signal process and do not perform a background subtraction. However, we do carry through a background
estimation procedure using the sidebands in data 
and compare the results with those obtained by applying the
procedure to the full \geant ~simulation.

The signal region is defined as a rectangle in the plane defined by two 
kinematic variables, the beam-energy substituted mass of the $B$ candidate \mes,
and the energy difference $\Delta E$~\cite{bib:babar}, where
\begin{eqnarray*}
m_{ES}&=&\sqrt{\left(\frac{\sqrt{s}}{2}\right)^2 - \left(\sum_{\alpha=1}^{n} {\bf p}^\ast_{\alpha}\right)^2 }, \\ 
\Delta E&=& \sum_{\alpha=1}^n \sqrt{ m_{\alpha}^2 + |{\bf p}^\ast_{\alpha}|^2 }-\sqrt{s}/2. 
\end{eqnarray*}
The index $\alpha$ is over the particles that make up the candidate 
$B$ meson system.  $m_{\alpha}$ are the masses of the particles and
${\bf p}^\ast_{\alpha}$ are their momenta measured 
in the \FourS\ center of mass frame.
$\sqrt{s}$/2 is one half
of the center of mass energy.
It is important to distinguish between the laboratory
frame, in which the beams have energies $E_{e^-}\approx9.0$\gev\ and 
$E_{e^+}\approx3.1$\gev, and the center of mass frame, in which the beam energies are equal
and the $\Upsilon(4S)$ is produced at rest.
In a symmetric-energy machine, these frames are identical, but at \pep2 the center of mass
frame is moving with respect to the laboratory with a boost factor $\beta\gamma\approx0.56$.

\section{Event selection}

We select events with at least five good quality tracks, with 20 or more hits in the drift chamber and originating from the beam spot.  Initially, we also require either two loosely identified,
oppositely-charged electrons, muons, or an electron-muon pair (for combinatoric background studies). In addition, 
we require $R_2< 0.8$, where $R_2$ is the ratio of the second and zeroth Fox-Wolfram moments~\cite{bib:FoxWolfram}. This requirement provides a first suppression of continuum events, which have a more collimated (``jet-like'') event topology than \BB\ events. At this stage, $R_2$ is evaluated in the center of mass frame using charged tracks only.
Events passing these loose event selection criteria are further selected by
requiring the electron and muon momenta in the lab frame to satisfy $p_e> 0.5\ {\rm GeV}/c$ and
$p_{\mu}> 1.0\ {\rm GeV}/c$. Electrons are identified on the basis of several quantities, primarily
$E/p_e$, where $E$ is the energy measured in the CsI electromagnetic calorimeter and $p_e$ is the
momentum determined by the tracking system. This ratio is near one for electrons or positrons.
Muons are identified primarily by the number of interaction
lengths of iron penetrated by the charged track through the magnet flux return, which is instrumented 
with layers of resistive-plate chambers. 
The events are then required to lie within a large, rectangular 
region in the $\Delta E$ vs. \mes\ plane: $m_{ES}> 5.0$~\gevcc and
$|\Delta E|< 0.8$~\gev.

After these basic event selection criteria, we apply a tighter set of particle identification requirements.  In addition, electrons and positrons are required to pass the conversions veto, which suppresses the leptons that are likely to have come from photon conversions in the material.  The kaon identification is based on combining the energy loss information (\dedx) from the silicon vertex detector and the drift chamber for momenta $p$ $<$ 0.6 GeV/c, with the knowledge of the Cherenkov angle from the DIRC for $p$ $>$ 0.6 GeV/c.  For the $B\rightarrow K\ell^+\ell^-$ modes we require that the kaon have $p_{K}$ $>$ 0.6 GeV/c in the lab frame.   

For the  $B^0\rightarrow K^{\ast0} \ell^+\ell^-$ channels, we reconstruct the $K^{\ast0}$ in the $K^+ \pi^-$ final state.  The kaon candidate is required to be identified as a kaon, while there are no particle identification requirements on the pion candidate.  The mass of the $K^+\pi^-$ pair is required to be within 75$~\mevcc$ of the $K^{\ast0}$ mass.  

\begin{figure}[!tb]
\begin{center}
\includegraphics[]{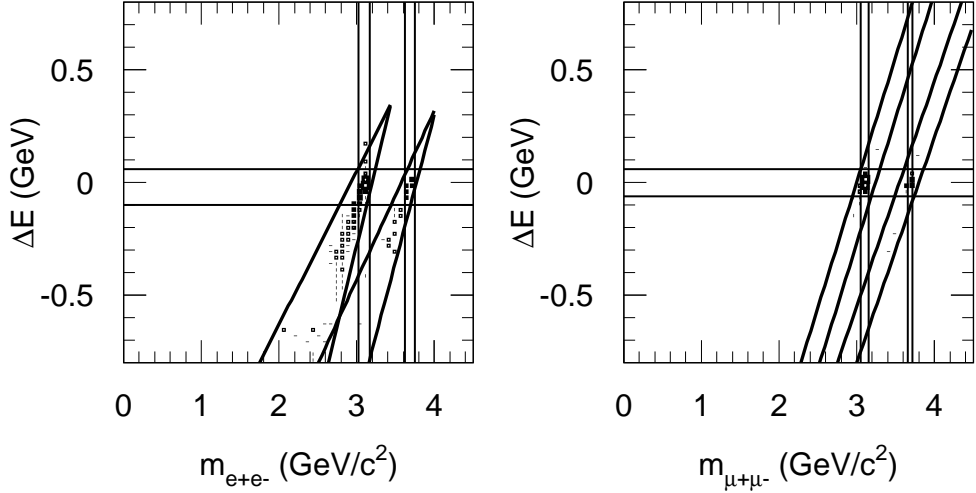}
\caption{Regions in the $\Delta E$ vs. $M_{\ell^+\ell^-}$ plane used to exclude $J/\psi \ K^{(\ast)}$ and $\psitwos \ K^{(\ast)}$ events: (a) shows the veto regions for the $J/\psi K^{(\ast)}$ and $\psitwos K^{(\ast)}$ events, where the $J/\psi$ or $\psitwos \rightarrow e^+ e^-$, (b) shows the veto regions for the $J/\psi K^{(\ast)}$ and $\psitwos K^{(\ast)}$ events, where the $J/\psi$ or $\psitwos \rightarrow \mu^+ \mu^-$.  The regions between the close pairs of vertical lines correspond to the nominal \jpsi\ and \psitwos\ resonance regions and are vetoed.
The diagonal lines veto the events with bremsstrahlung and track mismeasurement.  In the electron channel, the effects of bremsstrahlung are very substantial,
which necessitates the use of triangular selection criteria.  The signal region lies between the two horizontal lines except for the area excluded by the other vetos.
}
\label{fig:psiKveto}
\end{center}
\end{figure}

The next step is to veto $B\to \jpsi K^{(*)}$ and $B\to \psitwos K^{(*)}$. 
The requirements used for this purpose are somewhat complicated and are shown in Fig.~\ref{fig:psiKveto}. We remove events with dilepton masses
consistent with those of the \jpsi\ or \psitwos. These veto regions are shown as pairs of
vertical lines in the $\Delta E$ vs. $M_{\ell^+\ell^-}$ plane. However, bremsstrahlung or track mismeasurement
can result in a large departure of the dilepton mass from the \jpsi\ or
\psitwos\ masses.
Such departures are also accompanied by a shift in $\Delta E$. We, therefore, remove these events
by applying a correlated selection in this plane. Finally, a pair of horizontal lines is shown. There are veto
regions above and below these lines, while the signal region lies between them. This region between
the horizontal lines
corresponds to the 
requirement that $\Delta E$ be consistent with zero. This requirement is not part of the charmonium veto \emph{per se}
but is still useful to see on this figure, because it further restricts the allowed region in this 
plane. Veto regions for the dimuon final
state are defined in a similar way, but the definitions are somewhat different, since there is less bremsstrahlung.  
The $B \rightarrow J/\psi \ K$ modes can also pass this veto if the kaon is misidentified as a lepton (most often a muon).  In a similar way $B \rightarrow D^0 \pi$, where $D^0 \rightarrow K^- \pi^+$, can pass our selection criteria if both of the leptons are fake.  These effects can be suppressed by re-assigning the particle masses and excluding mass combinations around the ${J/\psi}$ and the ${D^0}$.  

Continuum background is suppressed by using a four-variable Fisher discriminant. 
The variables are the $R_2$, which at this stage is computed using both charged and neutral candidates, 
the cosine of the angle between the $B$ candidate and the $z$ axis 
in the center of mass frame, the cosine of the angle between 
the thrust axis of the signal and that of the rest of the event in the center of mass frame, and the invariant mass of the kaon and 
the oppositely charged lepton.
The last variable helps discriminate against the $D$ semileptonic decays, as it will tend to be below the $D$ mass.  Both continuum and \BB\ combinatorial background is further suppressed by requiring that the combined $B$ vertex probability be greater than 0.1\%.

Finally, we select the signal box in the $\Delta E$ vs. $m_{ES}$ plane: 5.272 $<$ $m_{ES}$ $<$ 5.286 GeV (3$\sigma$) and
$-0.10 < \Delta E < 0.06$~\gev\ ($|\Delta E| < 0.06$~\gev)
for the electron (muon) channels.   

\begin{figure}[!htb]
\begin{center}
\includegraphics[height=5.5in]{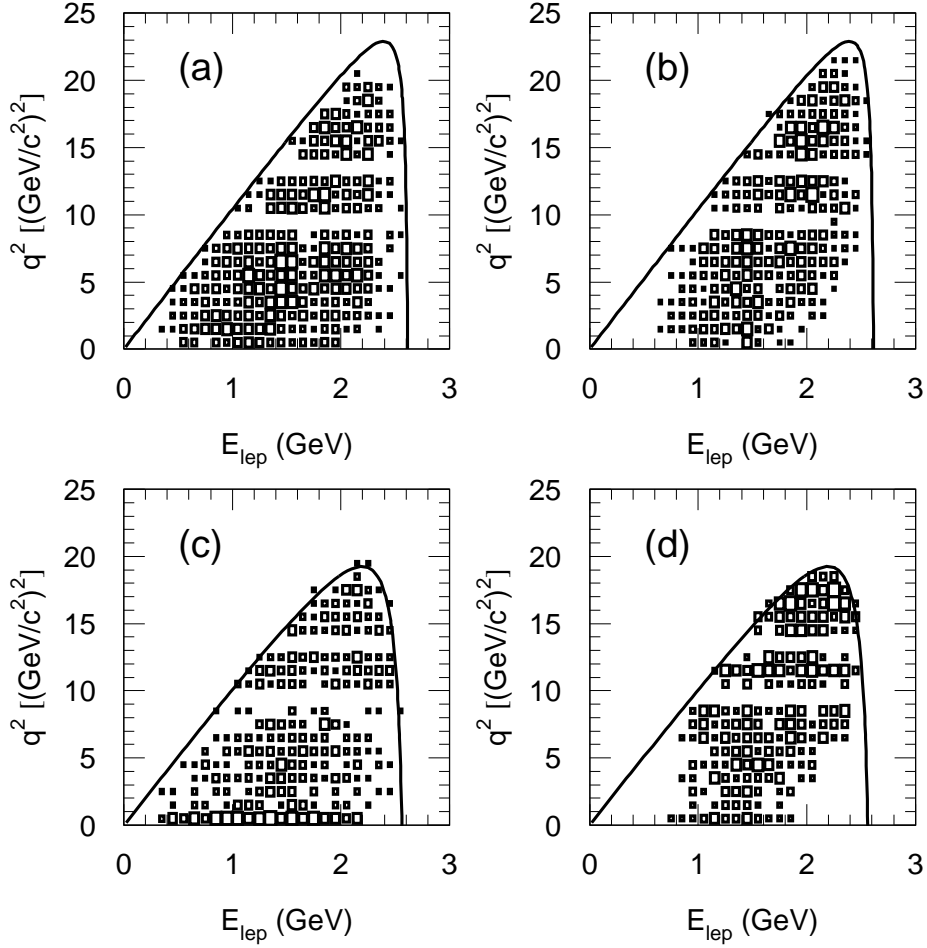}
\caption{Effects of the charmonium veto requirements on the dilepton mass distributions for (a) $B^+ \rightarrow K^+ e^+ e^-$, (b) $B^+ \rightarrow K^+ \mu^+ \mu^-$, (c) $B^0 \rightarrow K^{\ast0} e^+ e^-$, and (d) $B^0 \rightarrow K^{\ast0} \mu^+ \mu^-$.}
\label{fig:Dalitz_accept}
\end{center}
\end{figure}

The net effect of our analysis selection criteria on the Dalitz plot is shown in Fig.~\ref{fig:Dalitz_accept}.
The requirements to remove $B\to \jpsi K$ and $B\to \psitwos K$ decays are evident as
horizontal gaps, while those on the minimum lepton energy in the lab frame result in a
loss of acceptance around the edges of the Dalitz plot. 

\section{Physics results}
\label{sec:Physics}
%\emph{This section will contain the 4 plots of $\Delta E$ vs. $m_B$ for signal, as well as 4 $m_B$ projections.}

\begin{figure}[!tb]
\begin{center}
\includegraphics[height=5.7in]{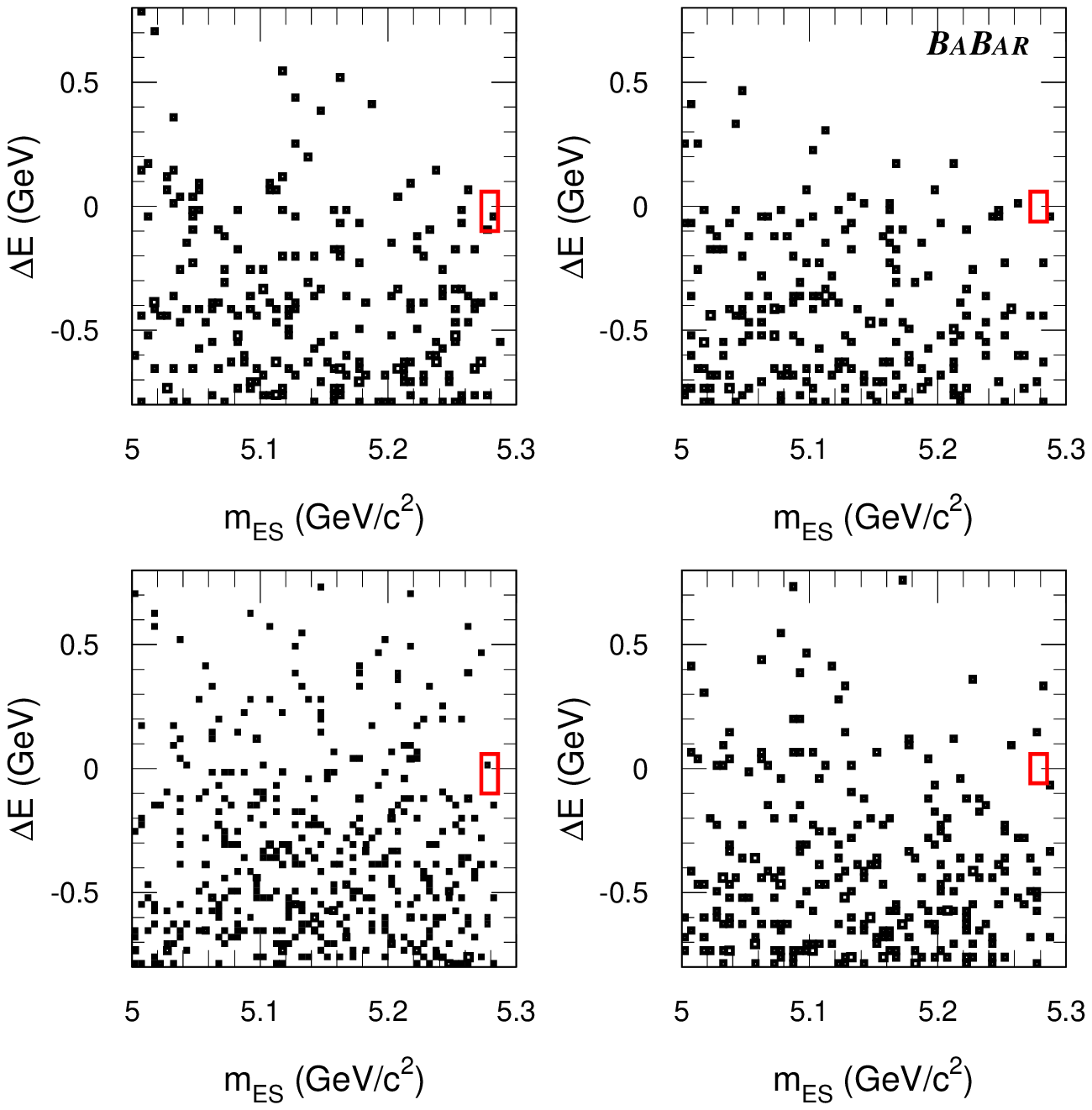}
\caption{$\Delta E$ vs. $m_{ES}$ (grand sideband) for data: (a) $\Bu\to \Kp e^+e^-$, (b) $\Bu\to \Kp \mu^+\mu^-$, (c) $\Bz\to \Kstarz e^+e^-$, and (d) $\Bz\to \Kstarz \mu^+\mu^-$.  The smaller boxes show the signal region.}
\label{fig:de_vs_Bmass_data}
\end{center}
\end{figure}

Figure~\ref{fig:de_vs_Bmass_data} shows a large $\Delta E$ vs.~$m_{ES}$ region (the ``grand sideband'') and a small box 
indicating the signal region for each of the four modes. 
%The $m_B$ range is the same for all final states: $5.272<m_B<5.286$ GeV. For
%$e^+e^-$ final states the $\Delta E$ range is $-0.10 < \Delta E < 0.06$ GeV, while for $\mu^+\mu^-$ final
%states it is $|\Delta E|<0.06$ GeV. 
These scatter plots show a trend of decreasing background as $\Delta E$
increases from negative to positive values; this is a characteristic of
\BB\ background, where the 
energy of three or four randomly selected tracks is not as high as the nominal $B$ energy in the center of mass frame.
Events from the continuum have a more uniform distribution in $\Delta E$ and are more important for positive
$\Delta E$ values. 

\begin{figure}[!tb]
\begin{center}
\includegraphics[height=5.5in]{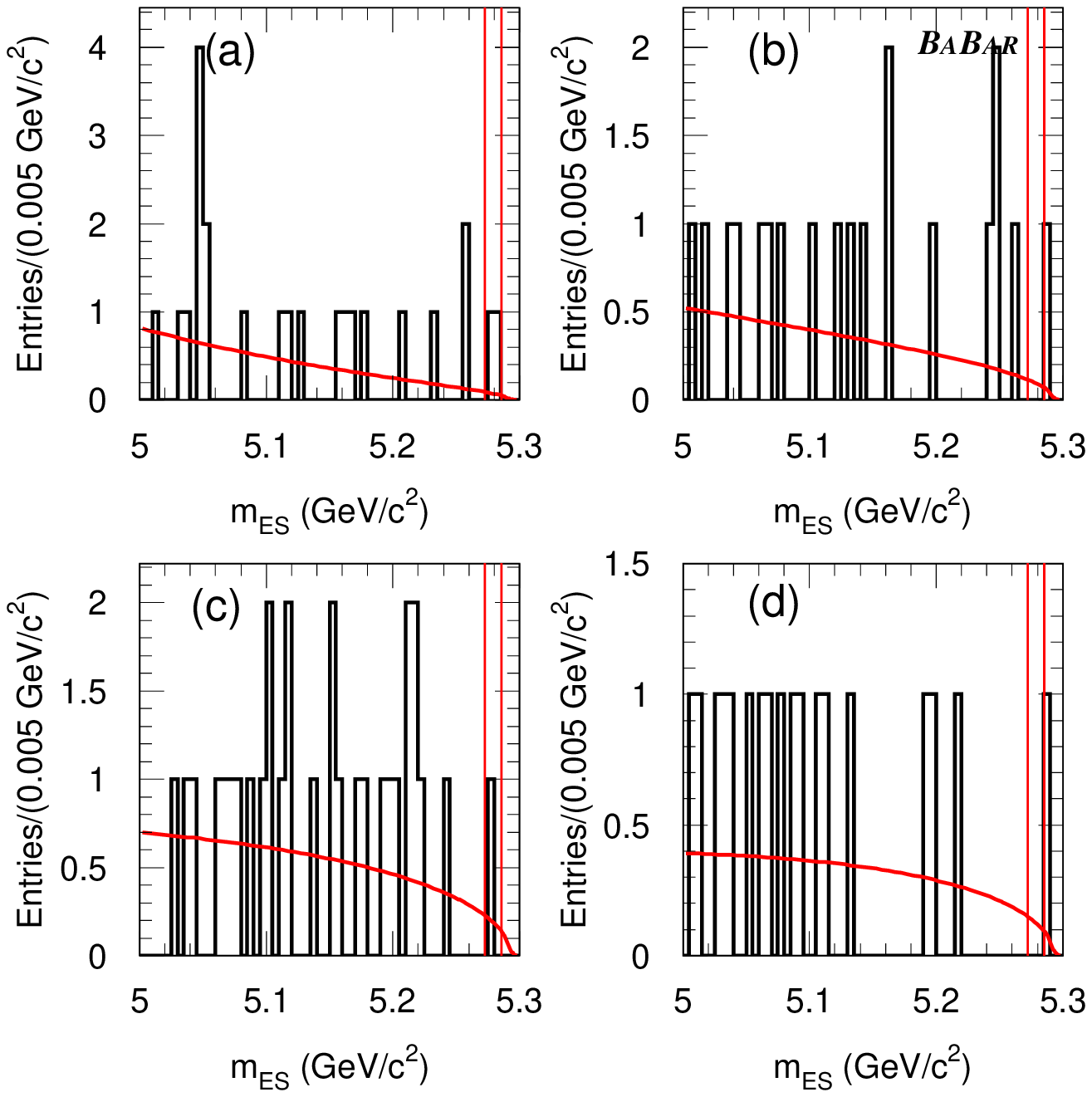}
\caption{$m_{ES}$ for data after all other event selection criteria are applied: (a) $\Bu\to \Kp e^+e^-$, (b) $\Bu\to \Kp \mu^+\mu^-$, (c) $\Bz\to \Kstarz e^+e^-$, and (d) $\Bz\to \Kstarz \mu^+\mu^-$.  The shape of the fit (the ARGUS function) is obtained from the large statistics sample of fast parametrized Monte Carlo events.  The lines indicate the signal region.}
\label{fig:bmass_data}
\end{center}
\end{figure}

Figure~\ref{fig:bmass_data} shows the data distribution in $m_{ES}$ after applying all other event selection criteria (including the requirement that the events lie within
the $\Delta E$ signal region). The distributions are fit to the ARGUS function~\cite{bib:argus}, whose shape is derived from the large statistics fast Monte Carlo 
simulation, and are used to estimate the background contribution to the $m_{ES}$ signal region. For this study,
however, we do not subtract this estimated background from the yield in the signal region; for the
purpose of setting the limits we assume that all events in the signal region might be due to the signal
events.

Table~\ref{tab:results} lists the signal efficiencies, total yield, 
the expected background, and the 90\% C.L. upper limits on the branching fractions.  The signal efficiencies were determined from the signal Monte Carlo events.  The efficiencies include the branching fractions for the \Kstarz\ modes.  The table also lists the total systematic error, determined as described in Section~\ref{sec:Systematics}.  It is important to note that because we decided before the measurement that our result would be expressed as an upper limit, we do not need to apply the Feldman-Cousins procedure (Ref.~\cite{bib:feld}), for a two-sided confidence interval, but instead use the original procedure specified in the Particle Data Book~\cite{bib:PDG}, in which zero observed events corresponds to an upper limit of 2.3 events.

\begin{table}[!htb]
\caption{Signal efficiencies, systematic uncertainties (combining the
uncertainties on the signal efficiencies and on the number of produced
$\Upsilon$(4S) mesons, as listed in Table~\ref{tab:syserr}), the number of
observed events, the number of estimated background events, and upper limits
on the branching fractions. In computing the upper limits we have assumed
$\BR(K^{*0}\to K^+\pi^-)=2/3$.
}
\vskip 0.5 cm 
\begin{center}
\begin{tabular}{|l|c|c|c|c|c|} 
\hline
Mode & Efficiency (\%) & Total systematic & Observed & Bkgd. estimated & $\BR/10^{-6}$ \\
     &                 & uncertainty (\%) & events   & from data          &  (90\% C.L.)\\
\hline\hline
$B^+ \rightarrow 
       K^+ e^+ e^-$             &  13.1 & 11.7 & 2 & 0.20  & $<$ 12.5  \\ 
$B^+ \rightarrow 
       K^+ \mu^+ \mu^-$         &  8.6  & 12.3 & 0 & 0.25  & $<$ 8.3  \\  
%$B^0 \rightarrow 
%       \Kstarz e^+ e^-$      &  ?  & 14.2 & 1 & 0.50  & $<$ 24.1  \\
%$B^0 \rightarrow 
%       \Kstarz \mu^+ \mu^-$  &  3.1  & 14.8  & 0 & 0.33 & $<$ 25.2  \\
$B^0 \rightarrow 
       \Kstarz e^+ e^-$       &  7.7  & 14.2 & 1 & 0.50  & $<$ 24.1  \\
$B^0 \rightarrow 
       \Kstarz \mu^+ \mu^-$   &  4.5  & 14.8  & 0 & 0.33 & $<$ 24.5  \\

\hline
\end{tabular}
\end{center}
\label{tab:results}
\end{table}

\section{Systematic studies}
\label{sec:Systematics}

 As we are setting a conservative limit by assuming that all
events observed in the signal region are in fact signal, the only systematic
effects that affect the limits are those due to the modeling 
of our signal efficiency and the number of produced $B$ mesons
in our sample. Table~\ref{tab:syserr} summarizes the systematic 
uncertainties that we have considered. The uncertainty
on the tracking efficiency is 2.5\% per track. Lepton and kaon identification
efficiencies are measured in control samples in data (e.g., a Bhabha sample for electrons), where a 
measure of the uncertainty is obtained by comparing different
control samples. We assign an error of 2.0\% per electron,
2.5\% per muon, and 3.0\% per kaon.  We have also studied the effects of 
momentum smearing and shifting on the $\Delta E$ selection criteria, and assign an error of
about 3.0\% for this effect.  Comparing data and Monte Carlo efficiencies
for the $m_{ES}$ requirement in the $J/\psi$ control region, we assign a 3.0\% error
for the modeling of this requirement.  We have compared the data and
Monte Carlo efficiencies for the $B$ vertex probability and the
Fisher discriminant selection criteria.  For the former, we assign
a systematic error of 3.0\% for the $B^+\to K^+\ell^+\ell^-$ modes and 4.0\% for the $B^0\to K^{\ast0}\ell^+\ell^-$ modes.  For the latter, we assign an error of 3.0\% for all
of the four modes.                                            
Furthermore, as a check that we understand our signal efficiencies, we
have compared the yields in data and Monte Carlo events
for the $B^+\to J/\psi K^+$ 
and $B^0\to J/\psi K^{*0}$ (with $\Kstarz\to K^+ \pi^-$) channels.
We have used the 
world average~\cite{bib:PDG} branching fractions for $B^+\to J/\psi K^+$ 
and $B^0\to J/\psi K^{*0}$ and found agreement within the statistical errors between the data and Monte Carlo samples.

\begin{table}[!htb]
\caption{Summary of the systematic uncertainties on the signal efficiencies
and the number of produced $\Upsilon$(4S) mesons as a percentage error on the
branching fraction, \BR. The total systematic error is the sum of the
individual contributions added in quadrature.}
\vskip 0.5 cm
\begin{center} 
\begin{tabular}{|c|c|c|c|c|} 
\hline
&\multicolumn{4}{c|}{$(\Delta \BR/\BR)$ (\%)} \\ \cline{2-5}
%&$(\Delta B/B)_{Kee}$,\% & $(\Delta B/B)_{K\mu\mu}$, \% & $(\Delta B/B)_{K^{*}ee}$,\%  & $(\Delta B/B)_{K^{*}\mu\mu}$,\% \\ \hline \hline
&${Kee}$ & ${K\mu\mu}$ & ${K^{*}ee}$ & ${K^{*}\mu\mu}$ \\ \hline \hline
Tracking efficiency         & 7.5  &  7.5  & 10.0 & 10.0     \\
Lepton identification       & 4.0  &  5.0  &  4.0 &  5.0     \\
Kaon identification         & 3.0  &  3.0  &  3.0 &  3.0     \\
$\Delta E$ requirement efficiency
                            & 2.0  &  2.5  &  3.3 &  1.5  \\
$m_{ES}$ requirement efficiency        & 3.0  &  3.0  &  3.0 &  3.0  \\
Vertex requirement efficiency       & 3.0  &  3.0  &  4.0 &  4.0  \\
Fisher requirement efficiency       & 3.0  &  3.0  &  3.0 &  3.0  \\
Number of produced $\Upsilon$(4S)  & 3.6  &  3.6  &  3.6 &  3.6  \\      
MC signal statistics        & 3.4  &  4.0  &  4.3 &  6.1  \\ 
\hline
Total systematic error      & 11.7 & 12.3  & 14.2 & 14.8  \\ 
 \hline
\hline
\end{tabular}
\label{tab:syserr}
\end{center}
\end{table}

\section{Summary}
\label{sec:Summary}
We have searched for rare $B$ decays $B \rightarrow K^{(\ast)} \ \ell^+ \ell^-$
in a sample of $3.7 \times 10^{6}$ \BB\ events.  We find no
observable signal for any of the four modes considered, and set 
preliminary 90\% C.L. upper limits on the branching fractions of
\begin{eqnarray*}    
\BR(B^+ \rightarrow K^+ e^+ e^-)  & <  & 12.5 \times 10^{-6}, \\
\BR(B^+ \rightarrow K^+ \mu^+ \mu^-) &  < & \phantom{2}8.3 \times 10^{-6}, \\
\BR(B^0 \rightarrow K^{\ast0} e^+ e^-) & < &  24.1  \times  10^{-6}, \\
\BR(B^0 \rightarrow K^{\ast0} \mu^+ \mu^-) & < & 24.5 \times 10^{-6}. \\
\end{eqnarray*}        
The limits for the $B^+\rightarrow K^+ \ell^+ \ell^-$ modes are comparable to those set by other experiments, while those for $B^0 \rightarrow K^{\ast0}\ell^+\ell^-$ are less sensitive with this data sample. We plan to analyze substantially more data in the near future.

\section{Acknowledgments}
\label{sec:Acknowledgments}

% Specific acknowledgments for this paper; remove if not needed.
The authors thank JoAnne Hewett and Gudrun Hiller for their advice on theoretical issues 
related to the analysis.

% Standard acknowledgments paragraph; must always be included.
We are grateful for the contributions of our \pep2\ colleagues in
achieving the excellent luminosity and machine conditions
that have made this work possible.
We acknowledge support from the
Natural Sciences and Engineering Research Council (Canada),
Institute of High Energy Physics (China),
Commissariat \`a l'Energie Atomique and
Institut National de Physique Nucl\'eaire et de Physique des Particules
(France),
Bundesministerium f\"ur Bildung und Forschung
(Germany),
Istituto Nazionale di Fisica Nucleare (Italy),
The Research Council of Norway,
Ministry of Science and Technology of the Russian Federation,
Particle Physics and Astronomy Research Council (United Kingdom), the
Department of Energy (US),
and the National Science Foundation (US). In addition, individual support 
has been received from the Swiss 
National Foundation, the A. P. Sloan Foundation, the Research Corporation,
and the Alexander von Humboldt Foundation.
The visiting groups wish to thank 
SLAC for the support and kind hospitality
extended to them.

\end{document}